\documentclass[%
 aip,
 amsmath,amssymb,
 reprint,%
]{revtex4-1}

\usepackage{graphicx,subcaption,adjustbox, float, flafter}
\usepackage{dcolumn}
\usepackage{bm}

\usepackage[utf8]{inputenc}
\usepackage[T1]{fontenc}
\usepackage{mathptmx,physics}
\usepackage{amsmath}
\usepackage{amssymb}
\usepackage{etoolbox}
\usepackage{float}
\usepackage{orcidlink}
\usepackage{fancyvrb}
\usepackage{color,hyperref}
\usepackage[capitalise]{cleveref}
\makeatletter
\def\@email#1#2{%
 \endgroup
 \patchcmd{\titleblock@produce}
  {\frontmatter@RRAPformat}
  {\frontmatter@RRAPformat{\produce@RRAP{*#1\href{mailto:#2}{#2}}}\frontmatter@RRAPformat}
  {}{}
}%
\makeatother
\newcommand{\mumaxp}{mumax$^+$}

\begin{document}


\title[Simulating altermagnets using \mumaxp{}]{Simulating altermagnets using \mumaxp{}}

\author{Lars Moreels}
\affiliation{DyNaMat, Department of Solid State Sciences, Ghent University, 9000 Ghent, Belgium}
\author{Nicolai Bechler}
\affiliation{Institute of Theoretical Solid State Physics, Karlsruhe Institute of Technology (KIT), 76049 Karlsruhe, Germany}
\author{Bartel Van Waeyenberge}
\author{Jonathan Leliaert}
\email{jonathan.leliaert@ugent.be}
\affiliation{DyNaMat, Department of Solid State Sciences, Ghent University, 9000 Ghent, Belgium}
\author{Jan Masell}
\email{jan.masell@kit.edu}
\affiliation{Institute of Theoretical Solid State Physics, Karlsruhe Institute of Technology (KIT), 76049 Karlsruhe, Germany}

\begin{abstract}
     In this paper, we demonstrate how altermagnets can be simulated in the recently released micromagnetic simulation package \mumaxp{}. We have added a new magnet class for $d$-wave altermagnets and demonstrate how \mumaxp{} is able to reproduce the analytical solutions for line profiles of the N\'eel vector and net magnetization for a Bloch domain wall. Next, we show simulation results of the magnon dispersion relation and its dependence on the anisotropic nature of the exchange interaction. Finally, we study the motion of a N\'eel skyrmion by applying a spin transfer torque. Altermagnets were implemented by extending the pre-existing code base for antiferromagnetic simulations. The object-oriented design of \mumaxp{} allows for a correct calculation of the magnetostatic field in multi-sublattice systems, a feature that many other micromagnetic simulators lack.
\end{abstract}

\maketitle

\section{Introduction}
Altermagnets~\cite{OG_ATM} have recently been identified as a new class of magnetic order, where the breaking of time-reversal symmetry is combined with compensated magnetic moments of antiferromagnetic order. This emerging research field in condensed matter physics has offered many promising results for the use of altermagnets in spintronic and magnonic applications~\cite{OG_ATM,spintronics, atm_app}. Numerical studies of altermagnets often utilize ab initio, atomistic DFT calculations~\cite{atomic_ATM,chiral_magnons,ortho_atm,hexa_atm,abinitio} or micromagnetic modeling of spin dynamics~\cite{Gomonay,spin_splitter,coupling_chiral_magnons} by exchange-coupling two ferromagnetic layers, representing two sublattices. However, when using the latter method, the magnetostatic interactions are calculated incorrectly. This is an important shortcoming, because in contrast to antiferromagnets, stray fields of altermagnetic textures are not always negligible~\cite{Gomonay,spin_splitter,multi_DW}. This demonstrates the need for updated micromagnetic tools to properly simulate altermagnets. In the current work, we present the implementation, verification, and usage of a new \texttt{Altermagnet} class in the micromagnetic simulation package \mumaxp{}~\cite{mumax+}, a recently released GPU-accelerated finite-difference solver. First, we outline the foundation and implementation of the added physics. Next, we verify its correctness by matching the simulation results to theoretical models for the static domain wall profile and the magnon dispersion relation. Finally, we simulate the motion of a N\'eel skyrmion under the influence of a spin transfer torque.

\section{Implementation and verification}

\subsection{The micromagnetic framework}

A micromagnetic theory assumes that the magnetization and related field quantities can be described by continuous, mesoscopic functions of space and time. In this sense, any microscopic crystal structure is coarse-grained, resulting in an averaged macrospin within a simulation cell. Multi-sublattice systems, such as antiferromagnets, are handled in \mumaxp{} by considering the sublattice magnetizations as coupled fields~\cite{mumax+,Giovanni} represented by multiple macrospins in each simulation cell.\\
The different interaction fields and related energy terms can be defined for each sublattice individually, as is described in Ref.~\cite{mumax+}. The total energy density relevant to the current work is given by
\begin{equation}
    \mathcal{E} = \mathcal{E}_\text{an} + \mathcal{E}_\text{demag} + \mathcal{E}_\text{exch} + \mathcal{E}_\text{hom} + \mathcal{E}_\text{inhom}\,.
\end{equation}
The first two terms describe the magnetocrystalline anisotropy field and the magnetostatic field respectively and are defined in Ref.~\cite{mumax+}, the term $\mathcal{E}_\text{exch}$ describes the exchange interaction within individual sublattices and the last two terms couple both sublattices via a homogeneous and an inhomogeneous exchange interaction. The three types of exchange are described in \cref{ssec:impl}.\\

Extensibility is the core design feature of  \mumaxp{}~\cite{mumax+}, meaning the package is built to facilitate easy addition of new physics in order to keep up with the quickly evolving research field of magnetism. By treating a ferromagnet as an instance from a \texttt{Ferromagnet} class, \mumaxp{} is able to couple multiple ferromagnets behind the scenes, effectively creating multi-sublattice systems, such as antiferromagnets, ferrimagnets and non-collinear antiferromagnets containing multiple spatially coinciding ferromagnetic sublattices. Exchange-coupled sublattice spins reside in the same simulation cell, allowing for a correct calculation of the magnetostatic field and the obsolescence of a small-angle approximation~\cite{mumax+}. This makes \mumaxp{} the right platform for the implementation of 
altermagnets.


\subsection{Implementation of 2D $d$-wave altermagnets in \mumaxp{}}
\label{ssec:impl}
 The implementation of altermagnets builds upon the one of antiferromagnets, and contains an additional anisotropic exchange field, which arises from a broken symmetry in the local environment of magnetic sublattice moments~\cite{Gomonay,Surfac_ATM, chiral_magnons}. The resulting spin order has $d$-wave symmetry in both real and reciprocal space~\cite{atomic_ATM}. In a micromagnetic framework, the anisotropic nature of the exchange interaction boils down to a different value of the exchange constant along the different principal axes of the simulation grid. In general, one can define the symmetric, positive definite matrix $\mathcal{A}^{(s)}$ for sublattice $s$ ($s=A,B$) to capture the spin stiffness~\cite{General_BC}
 \begin{equation}
     \mathcal{A}^{(A)} =
     \begin{pmatrix}
         A_1 & 0\\ 0 & A_2
     \end{pmatrix}\quad\,,\quad
     \mathcal{A}^{(B)} = 
     \begin{pmatrix}
         A_2 & 0\\ 0 & A_1
     \end{pmatrix}\,.
     \label{eq:diag_A}
 \end{equation}
 The $d$-wave symmetry of the altermagnetic spin ordering implies that the eigenbasis of $\mathcal{A}^{(s)}$ is rotated by $90^\circ$ between the two sublattices, thereby interchanging the roles of the eigenvalues $A_1$ and $A_2$, as is apparent in \cref{eq:diag_A}. This is done automatically in \mumaxp{} behind the scenes. In \cref{eq:diag_A}, the first (second) eigenvalue denotes the exchange stiffness constant along the $x$ ($y$) direction. The exchange field $\vb{H}_\text{ex}^{(s)}$ acting upon sublattice $s$ is then calculated with $\mathcal{A}^{(s)}_{ij}\partial_i\partial_j\vb{m}^{(s)}$, where $\vb{m}^{(s)}$ is the magnetization field of a single ferromagnetic sublattice $s$ normalized to the saturation magnetization, $\partial_i$ denotes a spatial partial derivative along the principal direction $i$ ($i,j=x,y$) of the simulation grid and a summation over repeated indices is implied. Explicitly, this leads to exchange fields given by
\begin{equation}
\begin{split}
    \vb{H}_\text{ex}^{(A)} &= \frac{2A_1}{\mu_0M_\text{S}}\pdv[2]{\vb{m}^{(A)}}{x} + \frac{2A_2}{\mu_0M_\text{S}}\pdv[2]{\vb{m}^{(A)}}{y}\,,\\
    \vb{H}_\text{ex}^{(B)} &= \frac{2A_2}{\mu_0M_\text{S}}\pdv[2]{\vb{m}^{(B)}}{x} + \frac{2A_1}{\mu_0M_\text{S}}\pdv[2]{\vb{m}^{(B)}}{y}\,,
\end{split}
    \label{eq:Hex_diag}
\end{equation}
for sublattice $A$ and $B$ respectively. This clearly illustrates the anisotropic nature of the exchange interaction, with different exchange strengths along different directions. In \cref{eq:Hex_diag}, $\mu_0$ is the vacuum permeability and $M_\text{S}$ is the saturation magnetization. The associated energy density is given by
\begin{equation}
\begin{split}
    \mathcal{E}_\text{exch}^{(A)} &= A_1 \left(\partial_x\vb{m}^{(A)}\right)^2 + A_2 \left(\partial_y\vb{m}^{(A)}\right)^2\\
    \mathcal{E}_\text{exch}^{(B)} &= A_2 \left(\partial_x\vb{m}^{(B)}\right)^2 + A_1 \left(\partial_y\vb{m}^{(B)}\right)^2\,.
\end{split}
\end{equation}
 An important observation is that the basis in which $\mathcal{A}^{(s)}$ is diagonal, does not necessarily coincide with the simulation grid directions. Since \mumaxp{} calculates derivatives along the latter, one can in general rotate the matrix $\mathcal{A}^{(s)}$ from its eigenbasis about an in-plane angle $\phi$ using a standard rotation matrix $\mathcal{R}(\phi)$. Explicitly, the matrix to be used in the calculation of $\vb{H}_\text{ex}^{(A)}$ is given by
\begin{align}
\hspace{-.41em}
\begin{split}
    \mathcal{A}^{(A)}_\phi &= \mathcal{R}(\phi)\mathcal{A}^{(A)}\mathcal{R}(\phi)^T\\ &= 
    \begin{pmatrix}
        A_1\cos^2\phi + A_2\sin^2\phi && \cos\phi\sin\phi\left(A_1 - A_2\right)\\
        \cos\phi\sin\phi\left(A_1 - A_2\right) && A_2\cos^2\phi + A_1\sin^2\phi
    \end{pmatrix}\,.
\end{split}
    \label{eq:rotated_A_tensor}
\end{align}
The expression for $\mathcal{A}^{(B)}$ can be obtained by interchanging $A_1$ and $A_2$. The full exchange field in a $d$-wave altermagnet acting upon sublattice $s$ is then calculated in \mumaxp{} with $\mathcal{A}^{(s)}_{\phi,{ij}}\partial_i\partial_j\vb{m}^{(s)}$ and the user is able to set the values of $A_1$, $A_2$ and $\phi$ as is shown in this minimal working example\footnote{The regular isotropic ferromagnetic exchange field can also be enabled in an Altermagnet instance. This can be used to create a 3D altermagnet, where the exchange in the out-of-plane direction is set by the regular exchange stiffness constant. The latter value should be subtracted from $A_1$ and $A_2$ to avoid a double exchange interaction in the in-plane directions.}.
\begin{Verbatim}[frame=single,fontsize=\small]
from mumaxplus import World, Grid, Altermagnet
import numpy as np

world  = World((1e-9, 1e-9, 1e-9))
grid   = Grid((32, 32, 1))
magnet = Altermagnet(world, grid)

magnet.alterex_1     = 15e-12
magnet.alterex_2     = 5e-12
magnet.alterex_angle = np.pi / 3
\end{Verbatim}
It should be noted that the matrix in~\cref{eq:rotated_A_tensor} is diagonal in every orthogonal basis if $A_1 = A_2$, as it reduces to the case of isotropic exchange interaction.\\
The calculation of the diagonal elements $\mathcal{A}^{(s)}_{\phi,ii}\partial_i\partial_i\vb{m}^{(s)}$ was readily available in \mumaxp{} and utilizes a second-order central difference scheme. The off-diagonal elements $\mathcal{A}^{(s)}_{\phi,ij}\partial_i\partial_j\vb{m}^{(s)}$ (with $i\neq j$) introduce mixed first-order derivatives. These were implemented using the second-order stencil
\begin{equation}
    \frac{\partial \vb{m}_{i,j}^{(s)}}{\partial x\partial y} \approx \frac{\vb{m}_{i+1, j+1}^{(s)} - \vb{m}_{i+1, j-1}^{(s)} - \vb{m}_{i-1, j+1}^{(s)} + \vb{m}_{i-1, j-1}^{(s)}}{4\Delta x\Delta y}\,,
\end{equation}
where the subscripts indicate cell coordinates and $\Delta x$ and $\Delta y$ denote the cell size in the $x$ and $y$ directions respectively.\\

Apart from the anisotropic exchange interaction, \mumaxp{} also contains two additional exchange fields to couple both sublattices~\cite{mumax+,Giovanni}. A first, homogeneous, exchange interaction couples spatially coinciding sublattice spins within the same simulation cell and is given by an energy density $\mathcal{E}_\text{hom} = -4\,A_0\left(\vb{m}^{(A)}\cdot\vb{m}^{(B)}\right)/a^2$ and is described by an exchange constant $A_0 < 0\,\mathrm{J/m}$ and a lattice parameter $a$ which can be set independently from the simulation cell size. A second, inhomogeneous, exchange interaction couples neighboring spins from different sublattices and is given by an energy density $\mathcal{E}_\text{inhom} = A_{12}\left(\nabla\vb{m}^{(A)}\right)\left(\nabla\vb{m}^{(B)}\right)$, with $A_{12} < 0\,\mathrm{J/m}$ an exchange constant similar to $A_1$ and $A_2$.\\

The remainder of this section is devoted to three test cases to verify the implementation of $d$-wave altermagnets in \mumaxp{}. The first example explores the static domain wall profile of the N\'eel vector and the net magnetization. A second test case investigates the magnon dispersion relation and its dependence on the angle $\phi$ between the exchange eigenbasis and the simulation grid. Finally, we study the motion of a N\'eel skyrmion by applying a spin transfer torque. In these sections, a detailed description of code snippets from the Python input script is given as a showcase of the use of \mumaxp{} in simulating $d$-wave altermagnets. The complete Python input scripts can be found in the supplementary material and can be run using version 1.2 of \mumaxp{}, which is freely available on Github (\href{https://github.com/mumax/plus}{https://github.com/mumax/plus}) under the GPLv3 license.

\begin{figure*}
    \centering
    \includegraphics[width=0.75\textwidth]{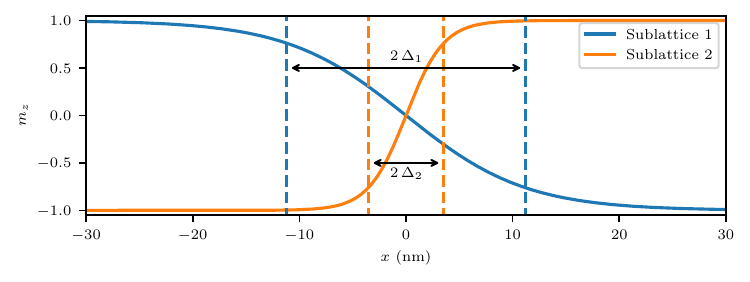}
    \caption{Domain wall profiles of the $z$-component of the sublattice magnetizations of an altermagnet with uncoupled sublattices. The double arrows and dashed lines indicate twice the domain wall widths $\Delta_1$ and $\Delta_2$ for sublattice 1 and 2 respectively, obtained from fitting the Walker solution \cref{eq:neel}. The exchange matrix eigenvalues are chosen as $A_1 = 50\,\mathrm{pJ/m} = 10\,A_2$ to clearly visualize the difference in the domain wall widths. Further details are explained in \cref{ssec:DW}.}
    \label{fig:DW_mags}
\end{figure*}

\subsection{Static domain wall profile}
\label{ssec:DW}
As discussed thoroughly in Ref.~\cite{Gomonay}, the anisotropic exchange stiffness leads to a non-vanishing net magnetization in a static domain wall. This can be understood intuitively as follows: we explained in \cref{ssec:impl} that the eigenvalues of the exchange matrix $\mathcal{A}$ are interchanged between both sublattices, meaning that when $A_1\neq A_2$ the exchange stiffness in some directions is different for each sublattice. This leads to a different domain wall width, as illustrated in \cref{fig:DW_mags} for the case of uncoupled sublattices. When the sublattices are exchange-coupled, the imperfect compensation of sublattice magnetization vectors leads to a non-vanishing net magnetic moment inside the domain wall.\\

Starting from a Lagrangian description, \citet{Gomonay} derived an expression for line profiles of the N\'eel vector $\vb{n}$ and the net magnetization $\vb{m}$ perpendicular to a domain wall separating two out-of-plane domains. The N\'eel vector follows the well known Walker solution in terms of the azimuthal angle $\varphi$ and the polar angle $\theta$~\cite{Gomonay,chiral_magnons_DW}:
\begin{equation}
    \theta = 2\arctan e^{p\xi}, \qquad \varphi = \text{const.}\,,
    \label{eq:neel}
\end{equation}
where $p=\pm1$ is the topological charge of the domain wall and $\xi = x/\Delta$ with $\Delta$ the domain wall width. The net magnetization is given by~\cite{Gomonay}
\begin{equation}
    \vb{m}(\xi) = C\left[\frac{\sinh^2\xi}{\cosh^3\xi}\left(\vb{e}_x\cos\varphi + \vb{e}_y\sin\varphi\right) + p\frac{\sinh\xi}{\cosh^3\xi}\vb{e}_z  \right]\,,
    \label{eq:net}
\end{equation}
where $\vb{e}_i$ ($i=x,y,z$) denote the principal directions along the simulation grid and $C$ is a constant depending on the magnetocrystalline anisotropy, the homogeneous exchange field (between sublattice spins in the same simulation cell) and the exchange matrix in \cref{eq:rotated_A_tensor}.\\

\begin{figure}
    \centering
    \includegraphics[width=0.8\linewidth]{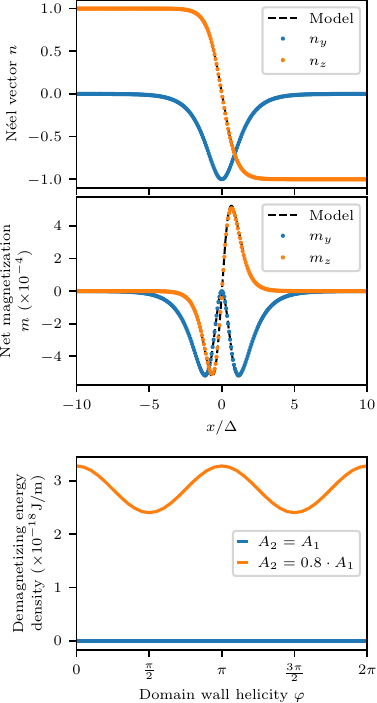}
    \caption{Profiles of the N\'eel vector (top) and the net magnetization (middle) of an altermagnet as modeled by \citet{Gomonay} (dashed line) and simulated using \mumaxp{} (points) for a Bloch domain wall (helicity $\varphi=-\pi/2$) with width $\Delta$. The bottom panel contains the magnetostatic energy per domain wall length calculated for domain walls with varying helicity $\varphi$ and exchange matrix eigenvalue $A_2$. Other simulation parameters are kept constant and are given in \cref{ssec:DW}.}
    \label{fig:DW_profile}
\end{figure}

To validate our implementation, we create an instance from the \texttt{Altermagnet} class with a thin film geometry. We also impose periodic boundaries in the (lateral) $y$-direction as to avoid any boundary effects that are not incorporated in the model.
\begin{Verbatim}[frame=single,fontsize=\small]
from mumaxplus import World, Grid, Altermagnet
import numpy as np

cs     = 0.5e-9
Nx     = int(256e-9 / cs)
Ny     = int(64e-9 / cs)

world  = World(cellsize        = (cs, cs, 5e-9), \
               pbc_repetitions = (0, 32, 0), \
               mastergrid      = Grid((0, Ny, 0))
               )

grid   = Grid((Nx, Ny, 1))
magnet = Altermagnet(world, grid)
\end{Verbatim}
The different material parameters in \mumaxp{} are defined as class properties~\cite{mumax+}. Consequently, they can be set for each created instance individually. Here we choose some prototypical values:
\begin{Verbatim}[frame=single,fontsize=\small]
magnet.msat  = 2.9e5
magnet.alpha = 0.01
magnet.ku1   = 2e5
magnet.anisU = (0, 0, 1)

# homogeneous sublattice coupling
magnet.afmex_cell = -5e-13

# inhomogeneous sublattice coupling
magnet.afmex_nn   = -2.5e-13

# exchange matrix eigenvalues
magnet.alterex_1 = 12.5e-12
magnet.alterex_2 = 10e-12

# exchange eigenbasis aligns with simulation grid
magnet.alterex_angle = 0 
\end{Verbatim}
Now we can create a two-domain state and let it relax to an energy minimum.
\begin{Verbatim}[frame=single,fontsize=\small]
dw = 10 # initial guess of wall width (in cells)

phi = - np.pi/2 # helicity of Bloch wall

m = np.zeros(magnet.sub1.magnetization.shape)

m[2, :, :,          0:Nx//2 - dw] = 1
m[0, :, :, Nx//2 - dw:Nx//2 + dw] = np.cos(phi)
m[1, :, :, Nx//2 - dw:Nx//2 + dw] = np.sin(phi)
m[2, :, :, Nx//2 + dw:          ] = -1

magnet.sub1.magnetization = m
magnet.sub2.magnetization = -m

magnet.minimize()
\end{Verbatim}
After minimizing, we extract the profiles of the N\'eel vector and the net magnetization and compare them to the theoretical solutions in Eqs (\ref{eq:neel}) and (\ref{eq:net}) as shown in \cref{fig:DW_profile}. Our simulation data agree very well with the theoretical Lagrangian model, with a maximal deviation of about $0.05\%$ and $3\%$ for the N\'eel vector (order of unity) and the net magnetization (order of $10^{-4}$) respectively.

The incomplete cancellation of sublattice spins inside the domain wall leads to a non-vanishing net magnetization, which results in a non-zero magnetostatic field. This highlights the shortcoming of other micromagnetic simulators in calculating this field, as the sublattice spins will be one simulation cell apart. In \mumaxp{}, however, they reside in the same point in the simulation space, resulting in a correct calculation of the magnetostatic field from a micromagnetic perspective. The bottom panel in \cref{fig:DW_profile} shows simulation data of the magnetostatic energy per domain wall length for a varying domain wall helicity $\varphi$, clearly illustrating the effect of an anisotropic exchange interaction on the magnetostatic field. These results show that a Bloch type domain wall minimizes the magnetostatic energy. The global off-set in the case where $A_1\neq A_2$ is expected due to the presence of surface charges. The order of magnitude of the energy density in \cref{fig:DW_profile} is also expected for a net magnetization $m=\norm{\vb{m}}$ of order $10^{-4\,\,}~\cite{coey}$. The data in the bottom panel of \cref{fig:DW_profile} highlights the importance of a correct calculation in altermagnetic textures with non-zero net magnetization.

\begin{figure*}
     \centering
    \begin{subfigure}[t]{0.49\linewidth}
        \includegraphics{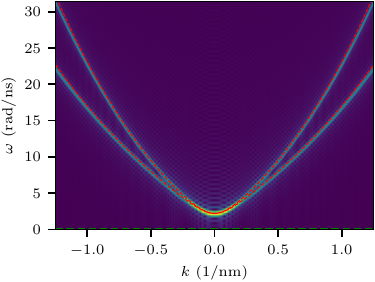}
        \label{fig:disp_0°}
    \end{subfigure}
    \begin{subfigure}[t]{0.49\linewidth}
        \includegraphics{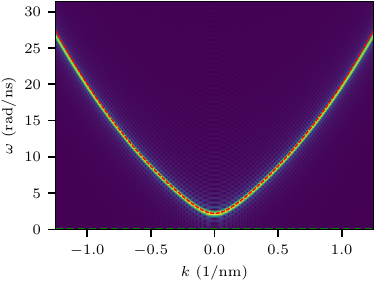}
        \label{fig:disp_45°}
    \end{subfigure}
     \caption{The dispersion relation of magnons in a 1D altermagnet as described in \cref{ssec:dispersion}. The angle $\phi$ between the exchange matrix eigenbasis and the simulation grid is set to $0^\circ$ in the left panel and $45^\circ$ in the right panel. The wave vector $\vb{k}$ lies along the length axis of the magnet. The dashed lines correspond to the model \cref{eq:disp}, while the intensity plot data is obtained with \mumaxp{}.}
     \label{fig:disp}
\end{figure*}
\subsection{Magnon dispersion relation}
\label{ssec:dispersion}

The anisotropic nature of the exchange interaction has its effect on propagation properties and dispersion of magnons. More precisely, magnons with large momenta $\vb{k}$ (small wavelengths) are mostly localized on individual sublattices~\cite{Gomonay} and therefore will experience two different exchange couplings in some directions, leading to a different effective magnon velocity. This lifts the degeneracy usually present in uniaxial antiferromagnets and creates two separate branches in the dispersion relation. \citet{Gomonay} derived an analytical expression for the magnon frequency $\omega$ in terms of the wave vector $\vb{k}$ starting from a spin Hamiltonian in reciprocal space. This result can also be obtained by linearizing the Landau-Lifshitz-Gilbert equation~\cite{LLG} and imposing a wave-like perturbation on the ground state. Using this procedure, we arrived at
\begin{equation}
    \omega_\pm = \pm\omega_\text{mag} \pm (\omega_\text{ext}+\omega_\text{alt})\,,
    \label{eq:disp}
\end{equation}
where the $+$ and $-$ sign correspond to the two different frequency branches. The first term represents the antiferromagnetic magnon frequency with an averaged exchange stiffness $A_\text{av} = \left(A_1+A_2\right)/2$, given by:
\begin{equation}
\begin{split}
    \omega_\text{mag} &= \frac{\gamma}{\mu_0 M_\text{S}}\sqrt{2K+\left(A_\text{av}-A_{12}\right)k^2}\\
    &\quad\times\sqrt{2K+\left(A_\text{av}+A_{12}\right)k^2-\frac{8A_0}{a^2}}\,,
\end{split}
\end{equation}
where $\gamma$ is the gyromagnetic ratio, $k$ the norm of the wave vector $\vb{k}$ and $K$ is the uniaxial anisotropy constant. An external field $H_\text{ext}$ lifts the degeneracy by causing branch splitting with a magnitude of $\omega_\text{ext} = \gamma\, H_\text{ext}$. The anisotropic character of altermagnetism causes an additional splitting, given by
\begin{equation}
\hspace{-.38em}
    \omega_\text{alt}= \frac{\gamma}{\mu_0M_\text{S}}\frac{A_1-A_2}{2}\left[\cos2\phi\left(k_x^2 - k_y^2\right) + 2\sin2\phi ~k_xk_y\right]\,.
    \label{eq:walt}
\end{equation}
Here, $k_x$ and $k_y$ are the components of the wave vector $\vb{k}$ along the principal directions of the simulation grid. The expression derived in Ref.~\cite{Gomonay} corresponds to \cref{eq:walt} for the case of $\phi=45^\circ$. In this case, magnons traveling along one of the principal axes of the simulation grid will experience an averaged exchange stiffness that is the same for both sublattices, which can also be seen by the diagonal elements in \cref{eq:rotated_A_tensor}. This leads to a degenerate dispersion and no branch splitting.\\

To further validate our implementation of $d$-wave altermagnets, we use the above dispersion as a second test case. Here, we simulate a 1D altermagnet using \mumaxp{}.
\begin{Verbatim}[frame=single,fontsize=\small]
from mumaxplus import Altermagnet, Grid, World
nx, ny, nz = 6000, 1, 1
cs = 0.5e-9

world  = World((cs, cs, cs))
magnet = Altermagnet(world, Grid((nx, ny, nz)))

magnet.sub1.magnetization = (0, 0, 1)
magnet.sub2.magnetization = (0, 0, -1)
\end{Verbatim}
We choose the same set of prototypical material parameters as in \cref{ssec:DW}, apart from the Gilbert damping constant $\alpha$: we decrease it to the value of $0.001$ to facilitate magnon propagation. The magnons are excited by an external field pulse with a sinc time profile in the middle of the simulation domain.
\begin{Verbatim}[frame=single,fontsize=\small]
import numpy as np
fmax = 1e13     # maximum excitation frequency
T    = 20e-12   # total simulation time

Bt = lambda t: (1e2 * np.sinc(2 * f * (t-T/2)),0,0)

# Put signal at the center of the simulation box
mask = np.zeros(shape=(nz, ny, nx))
mask[:, :, nx // 2 - 1:nx // 2 + 1] = 1

for sub in magnet.sublattices:
    sub.bias_magnetic_field.add_time_term(Bt, mask)
\end{Verbatim}

\begin{figure}[h]
    \centering
    \begin{subfigure}[t]{0.49\textwidth}
        \makebox[0pt][l]{\textbf{(a)}\hspace{0em}}%
        \hspace{4.8em}%
        \adjustbox{valign=t}{\includegraphics{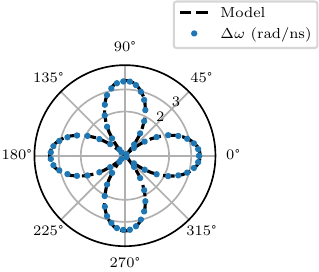}}
        \phantomsubcaption
        \label{fig:disp_freq}
    \end{subfigure}
    \begin{subfigure}[t]{0.49\textwidth}
        \makebox[0pt][l]{\textbf{(b)}\hspace{1em}}%
        \hspace{2em}%
        \adjustbox{valign=t}{\includegraphics{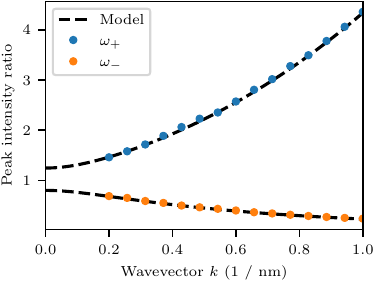}}
        \phantomsubcaption
        \label{fig:disp_int}
    \end{subfigure}
    \caption{Frequency gap $\Delta\omega$ at $k=0.75\,\mathrm{nm^{-1}}$ \textbf{(a)} and intensity ratio between both sublattices along the two different branches \textbf{(b)} of the magnon dispersion relation. The model (dashed lines) is described in \cref{ssec:dispersion} and Ref.~\cite{Gomonay}, while data points are obtained with \mumaxp{}. The angular coordinate in \textbf{(a)} corresponds to the relative angle $\phi$ between the exchange matrix eigenbasis and the simulation grid. Simulation details are described in \cref{ssec:dispersion}, but the total simulation time \texttt{T} is increased to $200\,\mathrm{ps}$ to improve the frequency resolution.}
    \label{fig:disp_verification}
\end{figure}

We save the sublattice magnetization states at a sampling rate of $2f_\text{max}$ and perform a spatiotemporal Fourier transform on this data so the results can be analyzed in reciprocal space $(\vb{k},\omega)$. The resulting dispersion relations are shown in \cref{fig:disp}, along with the model discussed above and derived in Ref.~\cite{Gomonay}. The deviation between the model and the simulation data for large values of $|\vb{k}|$ (small wavelengths) is to be expected, since a micromagnetic theory no longer forms an adequate description at such length scales. Our simulation data also confirm the prediction of mode degeneracy when $\phi=45^\circ$, as shown in the right panel of \cref{fig:disp}. \cref{fig:disp_freq} shows the frequency gap between both branches at a fixed value of $|\vb{k}|$ for varying angles $\phi$. \citet{Gomonay} also derived an expression for the relative magnon intensity between both sublattices along the two different branches. We calculated the same quantity, which is shown in \cref{fig:disp_int}. The simulation data obtained with \mumaxp{} agree very well with the model, with a maximal deviation of approximately $1.5\%$ and $2.5\%$ for the data in \cref{fig:disp_freq} and \cref{fig:disp_int} respectively.

\subsection{The skyrmion Hall effect}
\label{ssec:skh}
There have been recent reports of altermagnets exhibiting a skyrmion Hall effect induced by the anisotropic nature of the exchange interaction~\cite{spin_splitter,skh_1,skh_2,skh_3} when applying a Zhang-Li spin transfer torque~\cite{Zhang-Li}. While these models explicitly consider bilayer geometries, Jin et al.~\cite{skh_3} showed that the spin transfer torque in the limit of strong conduction electron coupling takes the typical Zhang-Li form with torques acting only within individual sublattices, similar to antiferromagnets~\cite{Yamane}.
\citet{Nicolai} showed that the skyrmion Hall effect originates from the truncation error induced by the finite-difference implementation of the spin transfer torque. In the case of a three point finite-difference stencil, the error and thus the apparent skyrmion Hall effect scale as $\mathcal{O}(c^2)$ with $c$ being the cell size. An accurate approximation of the continuum model may thus require numerical cell sizes smaller than the atomic lattice constant, which is a consequence of the numerical implementation and has no physical relevance. When utilizing a higher order finite-difference scheme instead, this effect is more strongly suppressed, as shown in Ref.~\cite{Nicolai}.\\

\begin{figure}[H]
    \centering
    \includegraphics{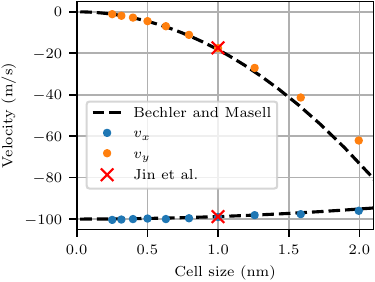}
    \caption{Skyrmion velocity components in function of the simulation cell size. The dashed lines represent the model in Ref.~\cite{Nicolai}, datapoints show simulated results using \mumaxp{}, and the crosses represent data taken from Ref.~\cite{skh_3}. Simulation details are given in \cref{ssec:skh}.}
    \label{fig:skh}
\end{figure}

We use the scaling relation derived in Ref.~\cite{Nicolai} as a third showcase of the altermagnet implementation in \mumaxp{}. We create an \texttt{Altermagnet} instance and initialize a N\'eel skyrmion.
\begin{Verbatim}[frame=single,fontsize=\small]
from mumaxplus.util import neelskyrmion as NSk

skyrmion = NSk(position     = magnet.center,
               radius       = 10e-9,
               charge       = -1,
               polarization = 1)
               
sub1, sub2 = magnet.sublattices
sub1.magnetization = skyrmion
sub2.magnetization = - sub1.magnetization()
magnet.minimize()
\end{Verbatim}
We choose material and excitation parameters as used in Ref.~\cite{skh_3} and monitor the position of the skyrmion in a time frame of $2\,\mathrm{ns}$. The calculated velocity is shown in \cref{fig:skh} for varying cell sizes. The model derived in Ref.~\cite{Nicolai} is also shown, confirming our altermagnet implementation in \mumaxp{}. The transverse skyrmion velocity decreases for decreasing cell sizes and its scaling is in agreement with the predictions in Ref.~\cite{Nicolai}.

\section{Conclusions}
We have outlined the implementation of a new magnet class in the \mumaxp{} simulation package to support the modeling of 2D $d$-wave altermagnets. This was done by the addition of an anisotropic exchange field term to the pre-existing code base designed to simulate antiferromagnets. The object-oriented design of \mumaxp{} allows for multiple sublattice spins to reside within the same simulation cell and, consequently, an accurate calculation of the magnetostatic field. This also entails that there is no need for a small-angle approximation between sublattice spins at the same point in space.\\
We have showcased this new feature by comparing the N\'eel vector and net magnetization profiles of a Bloch domain wall to an analytical model. Next, we derived and showed the magnon dispersion relation in a 1D altermagnet. Finally, we showed the cell size scaling of the skyrmion Hall effect. All three test cases showed excellent agreement between the results of \mumaxp{} and analytical models, thereby verifying its implementation. This also establishes the extensible nature of \mumaxp{}, where minimal effort was necessary in order to support $d$-wave altermagnets in a micromagnetic framework. This addition to the package is now able to nurture numerical studies in the emerging research field of altermagnetism. Further generalizations are to be expected in future release of \mumaxp{}, such as support for other exchange symmetries, for example $g$-wave, and three dimensional altermagnetic simulations.

\section{Supplementary Material}
The supplementary material contains the complete Python input scripts that can be used to generate the data from the paper. These can be run using v1.2 of \mumaxp{}, which can be found freely on Github (\url{https://github.com/mumax/plus}) under the GPLv3 license.

\begin{acknowledgments}
N.B. and J.M. acknowledge funding from the Deutsche Forschungsgemeinschaft (DFG, German Research Foundation) under the Project No. 547968854. B.V.W. acknowledges financial support from the SHAPEme project (EOS ID 400077525) from the FWO and F.R.S.-FNRS under the Excellence of Science (EOS) program. We acknowledge the Ghent University special research fund (bof/baf/1y/2024/01/005 and bof/baf/1y/2025/01/017 [J.L.] and bof/baf/4y/2024/01/288 [B.V.W.]).
\end{acknowledgments}

\section*{Author declarations}
\subsection*{Conflict of Interest}
The authors have no conflicts to disclose.
\subsection*{Author Contributions}
\textbf{Lars Moreels}: Formal Analysis, Investigation, Methodology, Software, Validation, Visualization, Writing/Original Draft Preparation, Writing/Review \& Editing. \textbf{Nicolai Bechler}: Conceptualization, Formal Analysis, Investigation, Validation, Writing/Review \& Editing. \textbf{Bartel Van Waeyenberge}: Formal Analysis, Funding Acquisition, Investigation, Project Administration, Supervision, Validation, Writing/Review \& Editing.
\textbf{Jonathan Leliaert}: Formal Analysis, Funding Acquisition, Investigation, Supervision, Validation, Writing/Review \& Editing. \textbf{Jan Masell}: Conceptualization, Funding Acquisition, Investigation, Project Administration, Supervision, Validation, Writing/Review \& Editing.

\subsection*{Code availability}
The data that support the findings of this study are available within the article and its supplementary material. \mumaxp{} is openly available on GitHub at \href{https://github.com/mumax/plus}{https://github.com/mumax/plus}.

\bibliography{bibliography}

@article{mumax+,
author={Moreels, Lars
and Lateur, Ian
and De Gusem, Diego
and Mulkers, Jeroen
and Maes, Jonathan
and Milo{\v{s}}evi{\'{c}}, Milorad V.
and Leliaert, Jonathan
and Van Waeyenberge, Bartel},
title={mumax+: extensible GPU-accelerated micromagnetics and beyond},
journal={npj Computational Materials},
year={2026},
month={Feb},
day={05},
volume={12},
number={1},
pages={71},
issn={2057-3960},
doi={10.1038/s41524-025-01893-y},
url={https://doi.org/10.1038/s41524-025-01893-y}
}

@article{Gomonay,
  title = {Structure,  control,  and dynamics of altermagnetic textures},
  volume = {2},
  ISSN = {2948-2119},
  url = {http://dx.doi.org/10.1038/s44306-024-00042-3},
  DOI = {10.1038/s44306-024-00042-3},
  number = {1},
  journal = {npj Spintronics},
  publisher = {Springer Science and Business Media LLC},
  author = {Gomonay,  O. and Kravchuk,  V. P. and Jaeschke-Ubiergo,  R. and Yershov,  K. V. and Jungwirth,  T. and Šmejkal,  L. and Brink,  J. van den and Sinova,  J.},
  year = {2024},
  month = jul 
}

@article{General_BC,
  title = {New Boundary-Driven Twist States in Systems with Broken Spatial Inversion Symmetry},
  volume = {119},
  ISSN = {1079-7114},
  url = {http://dx.doi.org/10.1103/PhysRevLett.119.127203},
  DOI = {10.1103/physrevlett.119.127203},
  number = {12},
  journal = {Physical Review Letters},
  publisher = {American Physical Society (APS)},
  author = {Hals,  Kjetil M. D. and Everschor-Sitte,  Karin},
  year = {2017},
  month = sep 
}

@article{chiral_magnons_DW,
  title = {Alignment-dependent gapless chiral split magnons in altermagnetic domain walls},
  author = {Zeng, Zhaozhuo and Jin, Zhejunyu and Yan, Peng},
  journal = {Phys. Rev. B},
  volume = {113},
  issue = {5},
  pages = {054401},
  numpages = {8},
  year = {2026},
  month = {Feb},
  publisher = {American Physical Society},
  doi = {10.1103/t6g4-gk34},
  url = {https://link.aps.org/doi/10.1103/t6g4-gk34}
}

@article{Surfac_ATM,
  title = {$d$-wave surface altermagnetism in centrosymmetric collinear antiferromagnets},
  author = {\ifmmode \mbox{\c{S}}\else \c{S}\fi{}a\ifmmode \mbox{\c{s}}\else \c{s}\fi{}\ifmmode \imath \else \i \fi{}o\ifmmode \breve{g}\else \u{g}\fi{}lu, Ersoy and Mertig, Ingrid and Lounis, Samir},
  journal = {Phys. Rev. B},
  volume = {114},
  issue = {2},
  pages = {L020406},
  numpages = {7},
  year = {2026},
  month = {Jul},
  publisher = {American Physical Society},
  doi = {10.1103/mkk5-b53m},
  url = {https://link.aps.org/doi/10.1103/mkk5-b53m}
}

@article{chiral_magnons,
  title = {Chiral Magnons in Altermagnetic $\text{RuO}_2$},
  volume = {131},
  ISSN = {1079-7114},
  url = {http://dx.doi.org/10.1103/PhysRevLett.131.256703},
  DOI = {10.1103/physrevlett.131.256703},
  number = {25},
  journal = {Physical Review Letters},
  publisher = {American Physical Society (APS)},
  author = {Šmejkal,  Libor and Marmodoro,  Alberto and Ahn,  Kyo-Hoon and González-Hernández,  Rafael and Turek,  Ilja and Mankovsky,  Sergiy and Ebert,  Hubert and D’Souza,  Sunil W. and Šipr,  Ondřej and Sinova,  Jairo and Jungwirth,  Tomáš},
  year = {2023},
  month = dec 
}

@misc{atomic_ATM,
Author = {Rodrigo Jaeschke-Ubiergo and Venkata-Krishna Bharadwaj and Warlley Campos and Ricardo Zarzuela and Nikolaos Biniskos and Rafael M. Fernandes and Tomas Jungwirth and Jairo Sinova and Libor Šmejkal},
Title = {Atomic Altermagnetism},
Year = {2025},
Eprint = {arXiv:2503.10797},
}

@article{OG_ATM,
  title = {Emerging Research Landscape of Altermagnetism},
  author = {\ifmmode \check{S}\else \v{S}\fi{}mejkal, Libor and Sinova, Jairo and Jungwirth, Tomas},
  journal = {Phys. Rev. X},
  volume = {12},
  issue = {4},
  pages = {040501},
  numpages = {27},
  year = {2022},
  month = {Dec},
  publisher = {American Physical Society},
  doi = {10.1103/PhysRevX.12.040501},
  url = {https://link.aps.org/doi/10.1103/PhysRevX.12.040501}
}

@article{spin_splitter,
  title = {Spin-Transfer Torque in Altermagnets with Magnetic Textures},
  author = {Vakili, Hamed and Schwartz, Edward and Kovalev, Alexey A.},
  journal = {Phys. Rev. Lett.},
  volume = {134},
  issue = {17},
  pages = {176401},
  numpages = {6},
  year = {2025},
  month = {Apr},
  publisher = {American Physical Society},
  doi = {10.1103/PhysRevLett.134.176401},
  url = {https://link.aps.org/doi/10.1103/PhysRevLett.134.176401}
}

@article{ortho_atm,
  title = {Altermagnetism in the orthorhombic $Pnma$ structure through group theory and DFT calculations},
  author = {Rooj, Suman and Saxena, Sugandha and Ganguli, Nirmal},
  journal = {Phys. Rev. B},
  volume = {111},
  issue = {1},
  pages = {014434},
  numpages = {11},
  year = {2025},
  month = {Jan},
  publisher = {American Physical Society},
  doi = {10.1103/PhysRevB.111.014434},
  url = {https://link.aps.org/doi/10.1103/PhysRevB.111.014434}
}

@Article{hexa_atm,
AUTHOR = {Chernov, Evgenii D. and Lukoyanov, Alexey V.},
TITLE = {Electronic Correlations in Altermagnet MnTe in Hexagonal Crystal Structure},
JOURNAL = {Materials},
VOLUME = {18},
YEAR = {2025},
NUMBER = {11},
ARTICLE-NUMBER = {2637},
URL = {https://www.mdpi.com/1996-1944/18/11/2637},
PubMedID = {40508635},
ISSN = {1996-1944},
DOI = {10.3390/ma18112637}
}

@article{abinitio,
  title = {Direct ab initio calculation of magnons in altermagnets: Method, spin-space symmetry aspects, and application to MnTe},
  author = {Sandratskii, L. M. and Carva, K. and Silkin, V. M.},
  journal = {Phys. Rev. B},
  volume = {111},
  issue = {18},
  pages = {184436},
  numpages = {16},
  year = {2025},
  month = {May},
  publisher = {American Physical Society},
  doi = {10.1103/PhysRevB.111.184436},
  url = {https://link.aps.org/doi/10.1103/PhysRevB.111.184436}
}

@article{multi_DW,
   title={Activation of anomalous Hall effect and orbital magnetization by domain walls in altermagnets},
   volume={112},
   ISSN={2469-9969},
   url={http://dx.doi.org/10.1103/vzmh-mxlz},
   DOI={10.1103/vzmh-mxlz},
   number={24},
   journal={Physical Review B},
   publisher={American Physical Society (APS)},
   author={Sorn, Sopheak and Mokrousov, Yuriy},
   year={2025},
   month=dec }

@article{coupling_chiral_magnons,
  title = {Strong Coupling of Chiral Magnons in Altermagnets},
  author = {Jin, Zhejunyu and Gong, Tianci and Liu, Jie and Yang, Huanhuan and Zeng, Zhaozhuo and Cao, Yunshan and Yan, Peng},
  journal = {Phys. Rev. Lett.},
  volume = {135},
  issue = {12},
  pages = {126702},
  numpages = {9},
  year = {2025},
  month = {Sep},
  publisher = {American Physical Society},
  doi = {10.1103/gn6c-1q19},
  url = {https://link.aps.org/doi/10.1103/gn6c-1q19}
}

@article{skh_1,
  title = {Altermagnetic skyrmions in two-dimensional lattices exhibiting the anisotropic skyrmion Hall effect},
  author = {Dou, Kaiying and He, Zhonglin and Du, Wenhui and Dai, Ying and Huang, Baibiao and Ma, Yandong},
  journal = {Phys. Rev. B},
  volume = {113},
  issue = {10},
  pages = {104402},
  numpages = {8},
  year = {2026},
  month = {Mar},
  publisher = {American Physical Society},
  doi = {10.1103/9vpq-hp7b},
  url = {https://link.aps.org/doi/10.1103/9vpq-hp7b}
}

@article{skh_2,
  title = {Dynamics and pinning for skyrmions in altermagnets},
  author = {Souza, J. C. Bellizotti and Reichhardt, C. J. O. and Saxena, A. and Reichhardt, C.},
  journal = {Phys. Rev. B},
  volume = {113},
  issue = {22},
  pages = {224424},
  numpages = {14},
  year = {2026},
  month = {Jun},
  publisher = {American Physical Society},
  doi = {10.1103/vplq-sd5k},
  url = {https://link.aps.org/doi/10.1103/vplq-sd5k}
}

@article{skh_3,
  title = {Skyrmion Hall Effect in Altermagnets},
  author = {Jin, Zhejunyu and Zeng, Zhaozhuo and Cao, Yunshan and Yan, Peng},
  journal = {Phys. Rev. Lett.},
  volume = {133},
  issue = {19},
  pages = {196701},
  numpages = {8},
  year = {2024},
  month = {Nov},
  publisher = {American Physical Society},
  doi = {10.1103/PhysRevLett.133.196701},
  url = {https://link.aps.org/doi/10.1103/PhysRevLett.133.196701}
}

@article{Zhang-Li,
  title = "{Roles of Nonequilibrium Conduction Electrons on the Magnetization Dynamics of Ferromagnets}",
  author = {Zhang, S. and Li, Z.},
  journal = {Physical Review Letters},
  volume = {93},
  issue = {12},
  pages = {127204},
  numpages = {4},
  year = {2004},
  month = {Sep},
  publisher = {American Physical Society},
  doi = {10.1103/PhysRevLett.93.127204}
}

@article{spintronics,
author={Jungwirth, T.
and Sinova, J.
and Wadley, P.
and Kriegner, D.
and Reichlov{\'a}, H.
and Krizek, F.
and Ohno, H.
and {\v{S}}mejkal, L.},
title={Altermagnetic spintronics},
journal={Nature Physics},
year={2026},
month={Jul},
day={01},
volume={22},
number={7},
pages={1012-1021},
issn={1745-2481},
doi={10.1038/s41567-026-03337-w},
url={https://doi.org/10.1038/s41567-026-03337-w}
}

@article{atm_app,
  title = {Altermagnets as a new class of functional materials},
  volume = {10},
  ISSN = {2058-8437},
  url = {http://dx.doi.org/10.1038/s41578-025-00779-1},
  DOI = {10.1038/s41578-025-00779-1},
  number = {6},
  journal = {Nature Reviews Materials},
  publisher = {Springer Science and Business Media LLC},
  author = {Song,  Cheng and Bai,  Hua and Zhou,  Zhiyuan and Han,  Lei and Reichlova,  Helena and Dil,  J. Hugo and Liu,  Junwei and Chen,  Xianzhe and Pan,  Feng},
  year = {2025},
  month = feb,
  pages = {473–485}
}

@INPROCEEDINGS{Nicolai,
  author={Bechler, Nicolai Timon and Masell, Jan},
  booktitle={2026 IEEE International Magnetic Conference - Short Papers (INTERMAG Short Papers)}, 
  title={Skyrmion Hall Effect in D-Wave Altermagnets}, 
  year={2026},
  pages={1-2},
  doi={10.1109/INTERMAGShortPapers68882.2026.11596479}}

@book{coey,
    title={Magnetism and Magnetic Materials},
    author={Coey, John M.D.},
    isbn={9780521816144},
    lccn={2010484052},
    series={Magnetism and Magnetic Materials},
    year={2010},
    publisher={Cambridge University Press},
    address = {Cambridge}
}

@article{Giovanni,
  title = "{Dynamics of domain-wall motion driven by spin-orbit torque in antiferromagnets}",
  author = {S\'anchez-Tejerina, Luis and Puliafito, Vito and Khalili Amiri, Pedram and Carpentieri, Mario and Finocchio, Giovanni},
  journal = {Physical Review B},
  volume = {101},
  issue = {1},
  pages = {014433},
  numpages = {10},
  year = {2020},
  month = {Jan},
  publisher = {American Physical Society},
  doi = {10.1103/PhysRevB.101.014433}
}

@article{LLG,
  title="{On the theory of the dispersion of magnetic permeability in ferromagnetic bodies. reproduced in collected papers of ld landau}",
  author={Landau, LD and Lifshitz, EM},
  journal={Pergamon, New York},
  year={1935}
}

@article{Yamane,
  title = {Spin-transfer torques in antiferromagnetic textures: Efficiency and quantification method},
  author = {Yamane, Yuta and Ieda, Jun'ichi and Sinova, Jairo},
  journal = {Phys. Rev. B},
  volume = {94},
  issue = {5},
  pages = {054409},
  numpages = {8},
  year = {2016},
  month = {Aug},
  publisher = {American Physical Society},
  doi = {10.1103/PhysRevB.94.054409},
  url = {https://link.aps.org/doi/10.1103/PhysRevB.94.054409}
}
\end{document}